\documentclass[preprint,nofootinbib,aps,superscriptaddress,eqsecnum]{revtex4-2} 
\usepackage{
graphicx,
hyperref,
amsmath,
amssymb,
charter,
xcolor,
ifluatex,
multirow}
 \usepackage{gensymb}       

\definecolor{Gray}{gray}{0.95}
\definecolor{RGray}{gray}{0.85}
\definecolor{CGray}{gray}{0.92}

\definecolor{tit}{rgb}{0.1,0.2,0.4}
\definecolor{blus}{cmyk}{1,1,0,0.6}
\definecolor{verde}{cmyk}{0.92,0,0.59,0.25}

\usepackage{tipa}
\usepackage{amsmath,amssymb,amsfonts, bm}
\usepackage{epsfig}
\usepackage{graphicx}
\usepackage{slashed}
\usepackage{color}

\usepackage{caption}
\usepackage{subcaption}
\usepackage{slashed} 
\usepackage{float}

\newcommand{\D}{{\cal D}}

\newcommand{\M}{{\cal M}}

\usepackage{calc}

\newcommand{\be}{\begin{equation}}
\newcommand{\ee}{\end{equation}}

\newcommand{\bea}{\begin{eqnarray}}
\newcommand{\eea}{\end{eqnarray}}

\newcommand{\bfig}{\begin{figure}}
\newcommand{\efig}{\end{figure}}

\makeatletter
\newcommand*{\rom}[1]{\expandafter\@slowromancap\romannumeral #1@}
\makeatother

\begin{document}
\title{Flavonic dark matter }
\author{Gauhar Abbas}
\email{email: gauhar.phy@iitbhu.ac.in}
\affiliation{ Department of Physics, Indian Institute of Technology (BHU), Varanasi 221005, India}
	
\author{Rathin Adhikari}
\email{rathin@ctp-jamia.res.in }
\affiliation{Centre for Theoretical Physics, Jamia Millia Islamia (Central University), New Delhi 110025, India}
	
\author{Eung Jin Chun}
\email{ejchun@kias.re.kr}
\affiliation{Korea Insitute for Advanced Study, Seoul 02455, Republic of Korea}
	
	\begin{abstract}
We first time show that a common solution to dark matter and the flavor problem of the standard model can be obtained in the framework of the  $\mathcal{Z}_{\rm N} \times \mathcal{Z}_{\rm M}$ flavor symmetry where the flavonic Goldstone boson of this flavor symmetry acts as a good dark matter candidate through the misalignment mechanism. Hierarchical mass pattern of quarks and charged leptons naturally follows from the discrete symmetry. For light active neutrinos, we construct the Dirac-type mass matrix which is preferred to fit the observed neutrino oscillation data with normal hierarchy.
Our model predicts the axion-like photon coupling characteristically different from the standard QCD axion,  and could be probed by the future X-ray or radio observations.
\end{abstract}


\maketitle
	\newpage

\section{Introduction}
The  Standard Model (SM) of particle physics is the most successful quantum theory of our universe providing a remarkable description of the elementary particles,  such as quarks and leptons  who constitute the matter of the universe,   and their interactions.    The SM,  notwithstanding with its triumph,  faces serious theoretical  imperfections and experimental failings.  In particular,  the discovery of the dark-matter (DM) is a dire experimental shortcoming of the SM.   On the theoretical side,  one of the critical problems is the so-called ``flavor-problem" of the SM.    The flavor problem is defined by the absence of any mechanism to explain the hierarchical structure of the masses of different flavors and their mixing in the SM.  The problem of neutrino masses and their oscillations can also be added to the flavor problem of the SM. This problem can be approached in different frameworks, such as, a technicolour framework  where the  vacuum-expectation-values   are sequential chiral condensates of an extended dark-technicolor sector providing a solution 
 \cite{Abbas:2017vws,Abbas:2020frs}, through an  Abelian flavor symmetry~\cite{Froggatt:1978nt,flavor_symm1,Chun:1996xv,flavor_symm2,flavor_symm3,Davidson:1983fy,Davidson:1987tr},  using loop-suppressed couplings to the Higgs~\cite{higgs_coup},  in a  wave-function localization scenario~\cite{wf_local},  through compositeness~\cite{partial_comp}, in an extra-dimension framework \cite{Fuentes-Martin:2022xnb}, and using   discrete symmetries\cite{Abbas:2018lga,Abbas:2022zfb,Higaki:2019ojq}.

It is remarkable to observe that  a particle-like  explanation to the problem of the dark matter,  and  a field theoretical solution to the flavor problem, such as the Frogatt-Nielsen (FN) mechanism \cite{Froggatt:1978nt}, are apparently mutually exclusive  and are poles apart.   
In the FN mechanism, the flavour problem is resolved  by an interaction of a new scalar field called flavon with the SM fermions  \cite{Froggatt:1978nt}:
\begin{equation} \label{FN}
 {\cal L}_{\rm Yuk}=  y_{ij} \left ( \chi \over \Lambda \right)^{n_{ij}} \bar{\psi}_{iL} H \psi'_{jR} +  \text{H.c.},
\end{equation} 
where $y_{ij}$ are order-one parameters and $\chi$ is the flavon field whose couplings (or the exponents $n_{ij}$) are controlled by continuous or discrete charges of the fields. 
After the flavor symmetry breaking,   the fermion Yukawa matrices are expressed in terms of the order parameter 
$\epsilon \equiv \langle \chi \rangle/\Lambda$.   Identifying the order parameter as the Cabibbo angle $\epsilon \approx 0.23$,  all the fermion masses and mixing matrices are
determined by powers of $\epsilon$.
Then the flavon is allowed to decay to the SM fermions at tree-level,  eliminating any possibility for this particle to be a DM candidate.  However, the axial degree of freedom of the flavon can be light enough to guarantee its stability. 
If the flavor symmetry is a continuous $U(1)$ symmetry, the axial flavon field could be identified with the QCD axion  \cite{Calibbi:2016hwq,Ema:2016ops,Bjorkeroth:2017tsz}
providing the solution to the strong CP problem as well as the axion dark matter \cite{Kim:2008hd}.
The flavor problem could be resolved by a discrete symmetry $\mathcal{Z}_{\rm N}$ allowing the flavon potential,
\begin{equation} \label{VN}
 V_{\mathcal{Z}_{\rm N}}= -\lambda {\chi^N\over \Lambda^{N-4}} + \text{H.c.}, 
\end{equation}
which is invariant under $\mathcal{Z}_{\rm N} $.
Upon the $\mathcal{Z}_{\rm N}$ breaking by the vacuum expectation value $\langle \chi \rangle = {v_F \over \sqrt{2}}$,   the flavonic Goldstone boson $\varphi$ receives the potential,
\begin{equation}
  V_{\mathcal{Z}_{\rm N}}= -{1\over4}|\lambda| {\epsilon^{N-4}} v_F^4  \cos\left( N {\varphi\over v_F}+ \alpha\right),
\end{equation}
where $\lambda=|\lambda|e^{i\alpha}$.  Thus, the axial flavon field can be very light for a sufficiently large $N$ and becomes a DM candidate whose abundance is generated by the misalignment mechanism \cite{misalign}. 

In this work,   we will set up a successful discrete flavor symmetry framework providing a solution of the flavor problem and,   show that the misalignment mechanism can generate the observed dark matter density in such a  framework.   We shall show  the axial flavon field can be a dark matter candidate associated with this discrete symmetry resolving the flavor problem,  and thus breaking the impasse posed by the demand of a joint solution of the DM and the flavor problem.   


\section{The $\mathcal{Z}_{\rm N} \times \mathcal{Z}_{\rm M}$ flavor symmetry  } 
\label{zn_zm}
The $\mathcal{Z}_{\rm N} \times \mathcal{Z}_{\rm M}$ flavor symmetry  is a new discrete symmetry product capable of providing a solution to the flavor problem of the SM through the FN mechanism \cite{Abbas:2018lga,Abbas:2022zfb}.   This was first proposed in reference \cite{Abbas:2018lga},  and later two prototypes of this symmetry are investigated in reference\cite{Abbas:2022zfb}.    In this work,  we use a $\mathcal{Z}_{\rm N} \times \mathcal{Z}_{\rm M}$ flavor symmetry that goes beyond the prototype symmetries discussed in reference \cite{Abbas:2022zfb}.  This is done by creating a flavor model where  the mass of the top quark does not originate from the tree level  SM  Yukawa operator.  This model is inspired by the hierarchical VEVs model \cite{Abbas:2017vws,Abbas:2020frs}, where  even the mass of the top quark arises from the dimension-5 operator.  This is done keeping in mind a possible technicolor origin of the $\mathcal{Z}_{\rm N} \times \mathcal{Z}_{\rm M}$ flavor symmetry.

Thus, we adopt the $\mathcal{Z}_{8} \times \mathcal{Z}_{22}$ flavor symmetry acting on the flavon field as well as the scalar and the fermionic sector of the SM as defined in table \ref{tab_z6z14}.  The generic form of the Lagrangian, after imposing the $\mathcal{Z}_{8} \times \mathcal{Z}_{22}$ flavor symmetry on the SM,  providing the masses to the SM fermions now reads,
\begin{eqnarray}
\label{mass1}
-{\mathcal{L}}_{\rm Yukawa} &=&          y_{ij}^u \bar{ \psi}_{L_i}^q  \tilde{H} \psi_{R_j}^{u} \left[  \dfrac{ \chi}{\Lambda} \right]^{n_{ij}^u} +         y_{ij}^d \bar{ \psi}_{L_i}^q  H \psi_{R_j}^{d} \left[  \dfrac{ \chi}{\Lambda} \right]^{n_{ij}^d} \nonumber \\
&+&            y_{ij}^\ell \bar{ \psi}_{L_i}^\ell  H \psi_{R_j}^{\ell} \left[  \dfrac{ \chi}{\Lambda} \right]^{n_{ij}^\ell} 
+  {\rm H.c.}, \\ \nonumber
&=&  Y^u_{ij} \bar{ \psi}_{L_i}^q  \tilde{H} \psi_{R_j}^{u}
+ Y^d_{ij} \bar{ \psi}_{L_i}^q  H \psi_{R_j}^{d}
+ Y^\ell_{ij} \bar{ \psi}_{L_i}^\ell  H \psi_{R_j}^{\ell}   + \text{H.c.},
\end{eqnarray}
where $i$ and $j$   represent family indices, $ \psi_{L}^q,  \psi_{L}^\ell    $ denote the quark and leptonic doublets, $ \psi_{R}^u,  \psi_{R}^d, \psi_{R}^\ell    $ are right-handed up, down type singlet quarks and  leptons, $H$ and $ \tilde{H}= -i \sigma_2 H^* $  denote the SM Higgs field and its conjugate and $\sigma_2$ is the second Pauli matrix.  We can write the  effective Yukawa couplings $Y_{ij}$  in terms of the expansion parameter  $   \epsilon= \dfrac{\langle \chi \rangle} { \Lambda}$ such that $Y_{ij} = y_{ij} \epsilon^{n_{ij}}$.

    \begin{table}
 \small
\begin{center}
\begin{tabular}{|c|c|c|c|c|c|c|c|c| c|c|c| c|c|c|}
  \hline
  Fields             &        $\mathcal{Z}_8$                    & $\mathcal{Z}_{22}$ & Fields             &        $\mathcal{Z}_8$                    & $\mathcal{Z}_{22}$   & Fields             &        $\mathcal{Z}_8$                    & $\mathcal{Z}_{22}$    & Fields             &        $\mathcal{Z}_8$                    & $\mathcal{Z}_{22}$     & Fields             &        $\mathcal{Z}_8$                    & $\mathcal{Z}_{22}$        \\
  \hline
  $u_{R}$                 &   $ \omega^2$  &$ \omega^{\prime 2}$        &$c_{R}$                 &   $ \omega^5$  & $ \omega^{\prime 5}$    &$t_{R}$                 &   $ \omega^6$  & $ \omega^{\prime 6}$       & $d_{R}$                 &   $ \omega^3$  &     $\omega^{\prime 3} $           & $s_{R}$                 &   $ \omega^4$  &     $\omega^{\prime 4} $           \\
  $b_{R}$                 &   $ \omega^4$  &     $\omega^{\prime 4} $     &   $\psi_{L,1}^q$                 &    $ \omega^2$  &    $\omega^{\prime 10} $      & $\psi_{L,2}^q$                 &  $ \omega$  &     $\omega^{\prime 9} $       &  $\psi_{L,3}^q$                 &    $\omega^{7} $  &      $\omega^{\prime 7} $ & $\psi_{L,1}^\ell$                 &   $ \omega^3$  &    $\omega^{\prime 3} $          \\
     $\psi_{L,2}^\ell$                  &   $ \omega^2$  &    $\omega^{\prime 2} $    &   $\psi_{L,3}^\ell$                 &   $ \omega^2$  &    $\omega^{\prime 2} $     &  $e_R$                 &   $\omega^{2} $  &     $\omega^{\prime 16} $          & $\mu_R$                 &  $\omega^5 $   &     $\omega^{\prime 19} $      &  $\tau_R $                 &   $ \omega^7$  &     $\omega^{\prime 21} $              \\
            $ \nu_{e_R} $                 &     $\omega^2 $    &     $1 $         & $   \nu_{\mu_R}$                 &     $\omega^5 $    &     $\omega^{\prime 3} $          &  $  \nu_{\tau_R} $                 &     $\omega^6 $    &     $\omega^{\prime 4} $        &   $\chi$                        & $ \omega$  &       $  \omega^\prime$       & $H$              &   1        &     1                  \\          
  \hline
     \end{tabular}
\end{center}
\caption{The charges of the SM  and the flavon fields under the $\mathcal{Z}_8 \times \mathcal{Z}_{22}$  symmetry,  where $\omega$ is the 8th,  and $\omega^\prime $ is the 22th root of unity. }
 \label{tab_z6z14}
\end{table}

The mass matrices of the  up and down-type quarks and charged leptons now can be written as,
\begin{align} \label{Mude}
\M_u & = \dfrac{v}{\sqrt{2}}
\begin{pmatrix}
y_{11}^u  \epsilon^8 &  y_{12}^u \epsilon^{5}  & y_{13}^u \epsilon^{4}    \\
y_{21}^u \epsilon^7     & y_{22}^u \epsilon^4  &  y_{23}^u \epsilon^{3}  \\
y_{31}^u  \epsilon^{5}    &  y_{32}^u  \epsilon^2     &  y_{33}^u  \epsilon 
\end{pmatrix}, 
\M_d   = \dfrac{v}{\sqrt{2}}
\begin{pmatrix}
y_{11}^d  \epsilon^7 &  y_{12}^d \epsilon^6 & y_{13}^d \epsilon^6   \\
y_{21}^d  \epsilon^6  & y_{22}^d \epsilon^5 &  y_{23}^d \epsilon^5  \\
 y_{31}^d \epsilon^4 &  y_{32}^d \epsilon^3   &  y_{33}^d \epsilon^3
\end{pmatrix}, 
\M_\ell =  \dfrac{v}{\sqrt{2}}
\begin{pmatrix}
y_{11}^\ell  \epsilon^9 &  y_{12}^\ell \epsilon^6  & y_{13}^\ell \epsilon^4   \\
y_{21}^\ell  \epsilon^{8}  & y_{22}^\ell \epsilon^5  &  y_{23}^\ell \epsilon^3  \\
 y_{31}^\ell \epsilon^{8}   &  y_{32}^\ell \epsilon^5   &  y_{33}^\ell \epsilon^3
\end{pmatrix}.
\end{align}

The masses of charged fermions are approximately  given by\cite{Rasin:1998je},
\begin{align}
\label{eqn51}
\{m_t, m_c, m_u\} &\simeq \{|y_{33}^u| \epsilon , ~ \left |y_{22}^u  - \frac {y_{23}^u y_{32}^u} {y_{33}^u  }   \right|  \epsilon^4 ,\\ \nonumber
&~ \left |y_{11}^u- \frac {y_{12}^u y_{21}^u}{y_{22}^u-y_{23}^u y_{32}^u/y_{33}^u}- \frac{y_{13}^u (y_{31}^u y_{22}^u-y_{21}^u y_{32}^u)-y_{31}^u y_{12}^u y_{23}^u}{(y_{22}^u- y_{23}^u y_{32}^u/y_{33}^u) y_{33}^u} \right| \epsilon^8\}v/\sqrt{2}  , \\ \nonumber
\{m_b, m_s, m_d\} & \simeq \{|y_{33}^d| \epsilon^3, ~ \left |y_{22}^d- \frac {y_{23}^d y_{32}^d} {y_{33}^d} \right| \epsilon^5,  \\ \nonumber
&  \left |y_{11}^d- \frac {y_{12}^d y_{21}^d}{y_{22}^d-y_{23}^d y_{32}^d/y_{33}^d}- \frac{y_{13}^d (y_{31}^d y_{22}^d-y_{21}^d y_{32}^d)-y_{31}^d y_{12}^d y_{23}^d}{(y_{22}^d- y_{23}^d y_{32}^d/y_{33}^d) y_{33}^d} \right| \epsilon^7\}v/\sqrt{2} ,  \\ \nonumber
\{m_\tau, m_\mu, m_e\} & \simeq \{|y_{33}^l| \epsilon^3, ~ \left|y_{22}^l- \frac {y_{23}^l y_{32}^l} {y_{33}^l} \right| \epsilon^5,\\ \nonumber 
& ~  \left |y_{11}^l- \frac {y_{12}^l y_{21}^l}{y_{22}^l-y_{23}^l y_{32}^l/y_{33}^l}- \frac{y_{13}^l \left( y_{31}^l y_{22}^l-y_{21}^l y_{32}^l \right) -y_{31}^l y_{12}^l y_{23}^l}{\left(  y_{22}^l- y_{23}^l y_{32}^l/y_{33}^l \right) y_{33}^l} \right| \epsilon^9\}v/\sqrt{2}.
\end{align}
The mixing angles of quarks read\cite{Rasin:1998je},
\begin{eqnarray}
\sin \theta_{12}  \simeq |V_{us}| &\simeq& \left|{y_{12}^d \over y_{22}^d}  -{y_{12}^u \over y_{22}^u}  \right| \epsilon, ~
\sin \theta_{23}  \simeq |V_{cb}| \simeq  \left|{y_{23}^d \over y_{33}^d}   -{y_{23}^u \over y_{33}^u}   \right| \epsilon^2,~
\sin \theta_{13}  \simeq |V_{ub}| \simeq  \left|{y_{13}^d \over y_{33}^d}    -{y_{12}^u y_{23}^d \over y_{22}^u y_{33}^d}      
- {y_{13}^u \over y_{33}^u}   \right|   \epsilon^3.
\end{eqnarray}

To obtain appropriate neutrino masses, we introduce three right handed  neutrinos $\nu_{eR}$, $\nu_{\mu R}$,$\nu_{\tau R}$  to the SM.  
We note that the Dirac mass operators for neutrinos, which conserve the total lepton number, can be written as
\begin{eqnarray}
\label{mass5}
-{\mathcal{L}}_{\rm Yukawa}^{\nu} &=&      y_{ij}^\nu \bar{ \psi}_{L_i}^\ell   \tilde{H}  \nu_{R_{j}} \left[  \dfrac{ \chi}{\Lambda} \right]^{n_{ij}^\nu} +  {\rm H.c.}. 
\end{eqnarray}

The Dirac mass matrix for neutrinos now reads,
\begin{equation}
\label{NM}
\M_{\D} = \dfrac{v}{\sqrt{2}}
\begin{pmatrix}
y_{11}^\nu  \epsilon^{25} &  y_{12}^\nu \epsilon^{22} & y_{13}^\nu   \epsilon^{21}  \\
y_{21}^\nu  \epsilon^{24}  & y_{22}^\nu \epsilon^{21} &  y_{23}^\nu   \epsilon^{20} \\
y_{31}^\nu \epsilon^{24}   &  y_{32}^\nu  \epsilon^{21}   &  y_{33}^\nu \epsilon^{20}
\end{pmatrix}.
\end{equation}

This mass matrix of the form (\ref{NM}) can lead naturally to the normal hierarchy masses given by
\begin{align}
\label{eqn5}
\{m_3, m_2,  m_1\} & \simeq \{|y_{33}^\nu| \epsilon^{20}, ~ \left|y_{22}^\nu- \frac {y_{23}^\nu y_{32}^\nu} {y_{33}^\nu} \right| \epsilon^{21},\\& ~  \left |y_{11}^\nu- \frac {y_{12}^\nu y_{21}^\nu}{y_{22}^\nu-y_{23}^\nu y_{32}^\nu/y_{33}^\nu}- \frac{y_{13}^\nu \left( y_{31}^\nu y_{22}^\nu-y_{21}^\nu y_{32}^\nu \right) -y_{31}^\nu y_{12}^\nu y_{23}^\nu}{ \left( y_{22}^\nu- y_{23}^\nu y_{32}^\nu/y_{33}^\nu \right) y_{33}^\nu} \right| \epsilon^{25}\}v/\sqrt{2}.\nonumber
\end{align}
From this, we can obtain the neutrino mass eigenvalues:
$\{m_3,m_2,m_1\} = \{0.05, 8.67 \times 10^{-3} , 1.73 \times 10^{-5} \}\, \text{eV}$
with the $y_{ij}^\nu$ couplings given in the appendix.   

The leptonic mixing angles are found to be,
\begin{eqnarray}
\label{numixing}
\sin \theta_{12}  &\simeq& \left|  {y_{12}^\ell \over y_{22}^\ell} -{y_{12}^\nu \over y_{22}^\nu}   \right| \epsilon, ~
\sin \theta_{23} \simeq  \left|  {y_{23}^\ell \over y_{33}^\ell} -  {y_{23}^\nu \over y_{33}^\nu}  \right|,~
\sin \theta_{13} \simeq \left| {y_{13}^\ell \over y_{33}^\ell}  - {y_{12}^\nu y_{23}^\ell\over y_{22}^\nu  y_{33}^\ell} -  {y_{13}^\nu \over y_{33}^\nu} \right|  \epsilon. 
\end{eqnarray}
From the above equation, we observe that the mixing angle  $\theta_{13}$ is of the order of the Cabibbo angle, and the mixing angle $\theta_{23} $ is of order one as expected from the structure of (\ref{NM}).  However, it leads to $\theta_{12} \propto \epsilon$ which is too small. Thus, one needs to rely on an unpleasant arrangement of the couplings $y^{l,\nu}_{i2}$ to fit the data.

We can investigate the inverted mass ordering as well.  For this purpose, we assign the following  charges to the right-handed neutrinos: $ \nu_{e_R}: \omega^6,  \omega^{\prime 4} $, $ \nu_{\mu_R}: \omega^6,  \omega^{\prime 4} $, $ \nu_{\tau_R}: \omega,  \omega^{\prime 21} $ under the $\mathcal{Z}_8 \times \mathcal{Z}_{22}$  symmetry.  This results the following mass matrix  of Dirac neutrinos,
\begin{equation}
\label{NMa}
\M_{\D} = \dfrac{v}{\sqrt{2}}
\begin{pmatrix}
y_{11}^\nu  \epsilon^{21} &  y_{12}^\nu \epsilon^{21} & y_{13}^\nu   \epsilon^{26}  \\
y_{21}^\nu  \epsilon^{20}  & y_{22}^\nu \epsilon^{20} &  y_{23}^\nu   \epsilon^{25} \\
y_{31}^\nu \epsilon^{20}   &  y_{32}^\nu  \epsilon^{20}   &  y_{33}^\nu \epsilon^{25}
\end{pmatrix}.
\end{equation}
The masses of neutrinos are approximately given by,
\begin{align}
\label{eqn5a}
\{m_3, m_2,  m_1\} & \simeq \{|y_{33}^\nu| \epsilon^{25}, ~ \left|y_{22}^\nu \right| \epsilon^{20},\left |y_{11}^\nu- \frac {y_{12}^\nu y_{21}^\nu}{y_{22}^\nu }  \right| \epsilon^{21}\}v/\sqrt{2}.
\end{align}
The neutrino mass eigenvalues are,
$\{m_3,m_2,m_1\} = \{1.70 \times 10^{-5}, 4.992 \times 10^{-2}, 4.92 \times 10^{-2} \}\, \text{eV}$
with the $y_{ij}^\nu$ couplings given in the appendix. 

The leptonic mixing angles turn out to be,
\begin{eqnarray}
\label{numixinga}
\sin \theta_{12}  &\simeq& \left|  {y_{12}^\ell \over y_{22}^\ell} -{y_{12}^\nu \over y_{22}^\nu}   \right| \epsilon, ~
\sin \theta_{23} \simeq  \left|  {y_{23}^\ell \over y_{33}^\ell} -  {y_{23}^\nu \over y_{33}^\nu}  \right|,~
\sin \theta_{13} \simeq \left| {y_{13}^\ell \over y_{33}^\ell}  - {y_{12}^\nu y_{23}^\ell\over y_{22}^\nu  y_{33}^\ell} -  {y_{13}^\nu \over y_{33}^\nu} \right|  \epsilon,
\end{eqnarray}
which are identical to that of the normal mass ordering.

Next we discuss the other possibilities of neutrino mass matrices in the model and their shortcomings. With the charge assignment for different fields as shown in Table- 1, we are allowed to write the pure Majorana mass operators for the left and right handed neutrinos.  
The  mass term ${\mathcal{L}}_{\rm Weinberg}^{\ell}$ in the Lagrangian   with left handed neutrino field, is given by the following Weinberg operator:
\begin{eqnarray}
\label{mass6}
-{\mathcal{L}}_{\rm Weinberg}^{\ell} &=&      h_{ij}^\nu  \frac{\bar{\tilde{\psi}}_{L_i}^{\ell }   H   \tilde{H}^\dagger \psi_{L_j}^{\ell }}{\Lambda}   \left[  \dfrac{ \chi^\dagger}{\Lambda} \right]^{n_{ij}^\nu}  +
{\rm H.c.},
\end{eqnarray}
where $\tilde{\psi}_{L_i}^{\ell } = i \sigma_2 \psi_{L_i}^c  $.

The above Lagrangian creates the following  neutrino mass matrix,
\begin{align}
\label{NM2}
\M_{L} & = \dfrac{v^2}{2 \Lambda }
\begin{pmatrix}
h_{11}^\nu  \epsilon^{24} &  h_{12}^\nu  \epsilon^{14} &  h_{13}^\nu  \epsilon^{14}  \\
 h_{12}^\nu  \epsilon^{14}  & h_{22}^\nu \epsilon^{4} &  h_{23}^\nu   \epsilon^{4} \\
 h_{13}^\nu  \epsilon^{14}   &  h_{23}^\nu  \epsilon^{4}   &  h_{33}^\nu \epsilon^{4}
\end{pmatrix}.
\end{align}
Let us note that $\Lambda \gg v$ in the realistic framework,  therefore the contribution of this mass matrix to neutrino masses is highly suppressed. 

We could have considered  type-\rom{1} seesaw mechanism \cite{seesaw}
for light neutrino masses. However, for that we have to introduce another new physics scale $\Lambda_1$ corresponding to the heavy right handed neutrino mass scale. This scale will not be related to the flavon field which is considered in this work. However, if  the right handed Majorana neutrino mass is related to the  scale $\Lambda$, the corresponding mass scale will not be heavy. This is because 
right handed neutrino mass operators would be written as $\mathcal{L}_{\rm M_R}$ given by
\bea
\mathcal{L}_{\rm M_R}  &=& c_{ij}  \chi \bar{\nu_{R_{i}}^c} \nu_{R_{j}} \left[  \frac{\chi }{\Lambda}  \right]^{n_{ij}^\nu}+
{\rm H.c.}.
\eea
Then the  right-handed Majorana mass matrix $\M_{R}$ is,
\begin{equation}
\label{MR}
\M_{R} =  \frac{v_F}{\sqrt{2}}
\begin{pmatrix}
c_{11} \epsilon^{26} & c_{12} \epsilon^{32}   & c_{13} \epsilon^{31} \\
  c_{12} \epsilon^{32}  & c_{22} \epsilon^{37} & c_{23}  \epsilon^{38}\\
c_{13} \epsilon^{31}  & c_{23} \epsilon^{36}  & c_{33} \epsilon^{35}
\end{pmatrix}.
\end{equation}
 for which the right handed neutrino mass scale is too small to be considered for Type I see-saw mechanism. 
So we refrain from considering see-saw mechanism for obtaining light neutrino mass.  Due to all above-mentioned points,
 the Dirac neutrinos with the mass matrices of the type (\ref{NM}) and   (\ref{NMa})  yielding the results (\ref{eqn5}),  (\ref{numixing}),  and   (\ref{eqn5a}),  (\ref{numixinga})
 are  preferred.

 Let us finally comment on the redundancy in constructing discrete flavor groups. One can find  different flavor symmetries reproducing the same flavor structures. For instance, we could have used a smaller flavor group like $\mathcal{Z}_{4} \times \mathcal{Z}_{17}$ to achieve what is  obtained in this section. Such a redundancy will be useful to predict different consequences in flavor violating processes and dark matter properties as will be discussed in the following sections.

\section{The axial flavon as cold dark matter}
\label{flav_dark}

In the framework of $\mathcal{Z}_{\rm N} \times \mathcal{Z}_{\rm M}$, the power of the flavon field in the flavon potential (\ref{VN}) is given by the least common multiple of $N$ and $M$ which we denote by $\tilde N$. 
Then the axial flavon mass is 
\begin{equation} 
\label{mphi1}
 m_\varphi^2={1\over8} |\lambda| \tilde N^2 \epsilon^{\tilde N-4} v_F^2.
\end{equation}
The axial flavon could be misaligned from the true vacuum during inflation and its initial amplitude sits at some point in the range $\varphi_0=(-\pi, +\pi) v_F/\tilde N$. Then, after the inflation, the boson field rolls down to the true vacuum to produce cold dark matter density of coherent oscillation.   Considering the linear approximation of the scalar potential, the axial boson field amplitude follows the equation of motion in the expanding universe: 
\begin{equation}
   \ddot{\varphi}+3H \dot{\varphi} + m_\varphi^2 \varphi \approx 0,
\end{equation}
which has the solution $\varphi(t)=  \varphi_0  2^{1\over4} J_{1\over4}(m_\varphi t)/(m_\varphi t)^{1\over 4}$. 
Its energy density  $\rho_\varphi = {1\over2}(\dot\varphi^2+m_\varphi^2 \varphi^2) $ at later time ($m_\varphi t\to \infty$) becomes $\rho_\varphi \approx m^2_\varphi \varphi_0^2 \sqrt{2}\Gamma(5/4)^2/\pi (m_\varphi t)^{3/2}$. Equating this with the dark matter density, $\rho_\varphi = 0.24\, {\rm eV}^4$ at the matter-radiation equality time $t_{eq}$, that is, $m_\varphi t_{eq} \approx 2 \times 10^{27} (m_\varphi/{\rm eV})$, we find the relation,
\begin{equation}
\label{mphi2}
  m_\varphi =3.4\times 10^{-3} {\rm eV} \left( 10^{12} {\rm GeV} \over \varphi_0 \right)^4,
\end{equation}
to get the right dark matter density.
Comparing this with (\ref{mphi1}) one finds the relation
\begin{equation}
\label{eq:v_F}
    v_F =2.5\times 10^7 \left( \frac{ \tilde N^{6} }{ a_0^8|\lambda| \epsilon^{\tilde N-4} } \right)^{1/10} {\rm GeV}
\end{equation}
taking $\varphi_0= a_0 v_F/\tilde N$. Thus, the required axial flavon mass is
\begin{equation}
\label{flav_mass}
    m_\varphi = 0.88 \times 10^{16} \left( \epsilon^{\tilde N-4} \tilde N^4 \frac{|\lambda|}{a_0^2} \right)^{2/5} {\rm eV}.
\end{equation}
For our flavor symmetry $\mathcal{Z}_{8} \times \mathcal{Z}_{22}$ discussed in the previous section, we have $\tilde N=88$ leading to  
\begin{align}
v_F  \approx 1.0 \times 10^{14}\, \rm{GeV}, ~~\text{and}~~
m_\varphi \approx   1.9 \times 10^{-3} \,  \rm{eV},
\end{align}
 considering $\epsilon=0.225$ with $|\lambda|=1$ and $a_0=1$.  Let us remark that one gets different values, $v_F  \approx 4.4 \times 10^{12}\, \rm{GeV}$ and $m_\varphi \approx  196 \,  \rm{eV}$, considering $\mathcal{Z}_{4} \times \mathcal{Z}_{17}$ with $\tilde N=68$ instead. 

For the longevity of the flavonic DM, its decay to electrons has to be forbidden, that is, $m_\varphi < 2 m_e$ which requires
\begin{equation} \label{Nmin}
\tilde N > 53, ~~\mbox{and} ~~ v_F > 4 \times 10^{11} \mbox{GeV}.
\end{equation}
From (\ref{mass5}) and (\ref{NM}), one can see that the largest coupling of the DM with neutrinos is $g_{\varphi \nu \nu} \sim 10 \sqrt{2} \epsilon^{20} v/v_F$ and thus the flavonic DM decay to neutrinos are highly suppressed.  For $\tilde{N} = 54 - 120$, we obtain the flavonic dark matter range $10^{-11}-10^6$eV.

\section{Phenomenology of flavonic dark matter}
\label{flav_couplings}
\label{decay_sign}

The axial degree of freedom $\varphi$  of the flavon field $\chi$ remains light and contribute to the flavor changing processes as studied for the falvorful axion model \cite{Bjorkeroth:2018dzu}.  The similar calculation can be made also for our case with the discrete flavor symmetry breaking. Let us first note that our discrete symmetry enforces an automatic $U(1)$ symmetry 
 in the Yukawa matrices (\ref{Mude})
under which the fermion fields $\psi^q_{L,i}$, $\psi^u_{R,i}$, $\psi^d_{R,i}$, $\psi^l_{L,i}$, and $\psi^l_{R,i}$ carry the following charges:
\begin{equation} \label{xf}
    x^q_i=(4,3,1),~ x^u_i=(-4,-1,0),~ x^d_i=(-3,-2,-2),~ x^l_i=(3,2,2),~\mbox{and}~x^e_i=(-6,-3,-1),
\end{equation}
 assigning the charge $+1$ to the order parameter $\epsilon$.
respectively for $i=1,2,3$.  Therefore, the field transformation of $\psi^f_{L/R,i} \to \exp(i x^f_i \varphi/v_F) \psi^f_{L/R,i}$ for $f=q,u,d,l,e$ will induce the derivative couplings of the axial boson:
\begin{equation}
   -{\cal L}_\varphi = \frac{\partial_\mu \varphi}{v_F} \sum_{f,i} x^f_i \,\bar{\psi}^f_{L/R,i} \gamma^\mu \psi^f_{L/R,i}.
\end{equation}
Then, the mass diagonalization of the quarks and leptons, performed by the diagonalization matrices $U_{u,d}$ ($V_{u,d}$) for the left-handed (right-handed) up and down quarks, and $U_e$ ($V_e$) for the left-handed (right-handed) charged leptons, will lead to the following FCNC couplings:
\begin{equation} \label{VA}
   - {\cal L}_\varphi = \frac{\partial_\mu \varphi}{v_F} \sum_{f=u,d,e} 
   \bar{f}_i \left( \gamma^\mu V^f_{ij} -\gamma^\mu\gamma_5 A^f_{ij}\right) f_j,  
\end{equation}
where $V^{f}/A^f=X_L^{f}\pm X_R^{f}$ with $X_L^{u,d} = U^\dagger_{u,d} x^q U_{u,d}$, $X_R^{u,d} = V^\dagger_{u,d} x^{u,d} V_{u,d}$, and $X_L^{e} = U^\dagger_{e} x^l U_{e}$, $X_R^{e} = V^\dagger_{e} x^{e} V_{e}$.

\begin{figure}[h]
    \centering
\includegraphics[width=10cm]{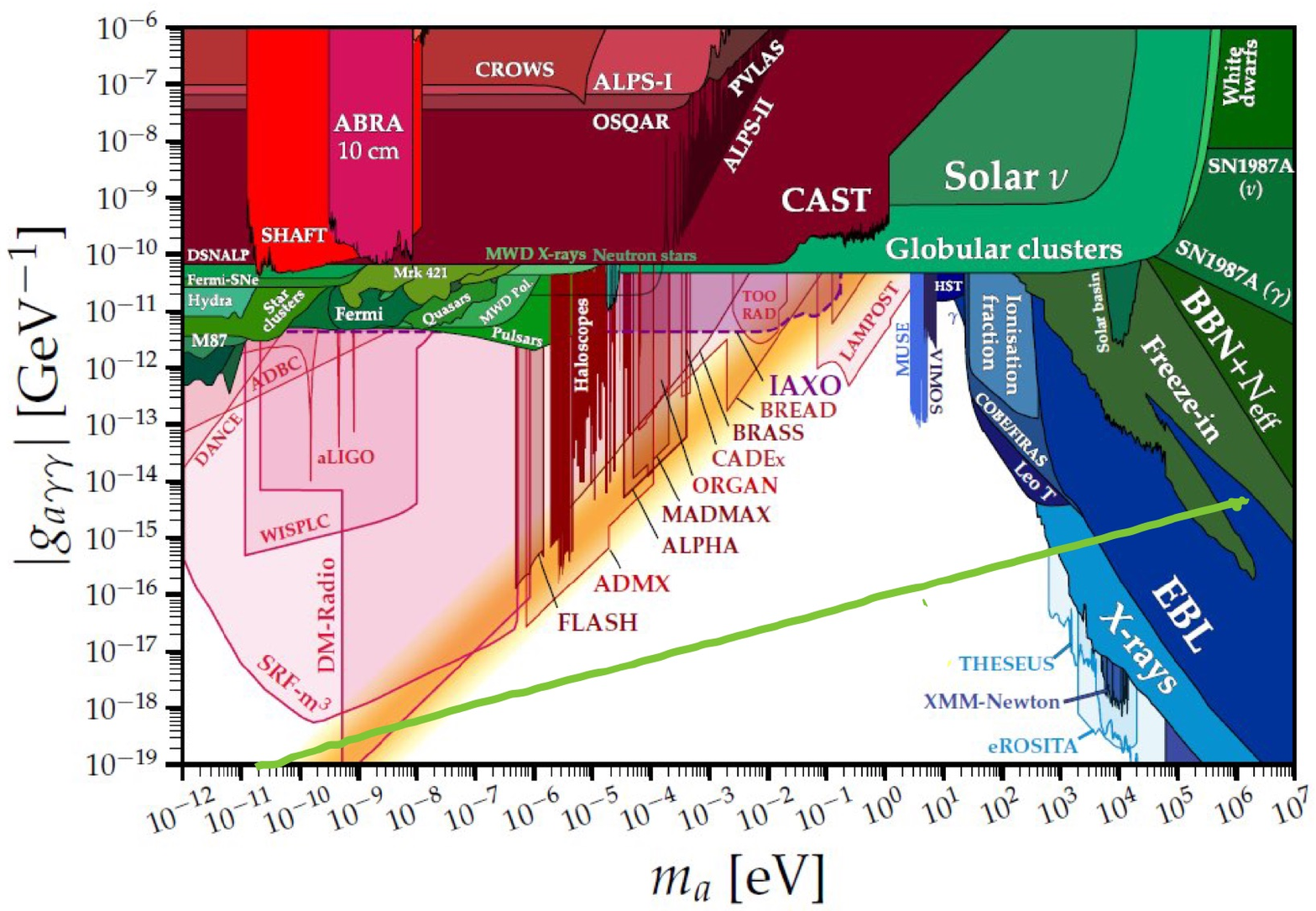} 
  \caption{ The prediction of flavonic dark matter (thick green line) and axion-like particle ($a\equiv \varphi$) searches \cite{Antel:2023hkf}. }
    \label{fig:fips}
\end{figure}

The most stringent bound on the flavon scale $v_F$ comes from the FCNC process $K^+ \to \pi^+ \varphi$ \cite{Bjorkeroth:2018dzu}:
\begin{eqnarray}
    v_F \gtrsim 7 \times 10^{11} V^d_{21} {\rm GeV}, 
\end{eqnarray}
where we have $V^d_{21} \approx \epsilon$.
Notice that this bound is trivially satisfied in our the flavonic DM scenario requiring (\ref{Nmin}). 
  The future sensitivity of the branching ratio of $K \rightarrow \pi \nu \bar{\nu}$ at NA62 is about $0.9 \times 10^{-10}$ and
the limit on $K\to \pi \varphi$ could be improved correspondingly, but only up to $v_F \sim 10^{12}$ GeV \cite{Bjorkeroth:2018dzu}.

The most promising channel to observe the axial flavon DM would be its coupling to photons 
\begin{align}
    \mathcal{L}^{\varphi \gamma \gamma}_{\rm eff} = \frac{1}{4} g_{\varphi\gamma\gamma} \varphi  F^{\mu\nu} \Tilde{F}_{\mu \nu},
\end{align}
which arises from the axial coupling of (\ref{VA}) leading to $g_{\varphi\gamma\gamma} = \frac{\alpha}{2 \pi v_F} \sum_{f,i}  N_{cf} A^f_{ii}  Q_f^2$ where $N_{cf}$ is the color factor of the fermion $f$.
For (\ref{xf}), we obtain $g_{\varphi\gamma\gamma} = \frac{\alpha}{2 \pi v_F} \frac{5}{3}$. 
Taking this relation with   (\ref{eq:v_F}) and (\ref{flav_mass}), we show in Fig. 1 the predicted photon coupling vs. the flavonic DM mass denoted by the thick green line which is overlaid in Fig.~15 of \cite{Antel:2023hkf} identifying the axion-like particle $a$ to our axial flavon $\varphi$. One can see that the DM mass larger than about 1 keV, corresponding to  $\tilde N < 67$ and $v_F < 4 \times 10^{12}$ GeV, is ruled out. This is also found from recent bound on $g_{\varphi\gamma\gamma}$ from INTEGRAL/SPI data \cite{INT}. Above KeV
mass range can be further examined by the forthcoming experiment THESEUS  \cite{Thorpe-Morgan:2020rwc}. Note also that our prediction overlaps with that of the GUT-scale QCD axion at around $10^{-9}$ eV which can be looked for in the future \cite{DMRadio:2022jfv}.

\section{Summary}
\label{sum}

The absence of any explanation to the discovery of DM is one of the most serious flaws in the framework of the SM.   
Furthermore,  the flavor structure of the SM is an challenging theoretical puzzle.    This problem is bizarre in the sense that the mass hierarchy among  the second and third generation quarks is very different from that of the first generation quarks.  Moreover,  the quark mixing is also entirely different from the neutrino mixing.  A solution of the flavor problem should not only  produce an explanation for the charged fermion masses and mixing,  it must account for the neutrino masses and mixing.  
 
A bosonic field called flavon may interact with the SM fermions to produce a hierarchical spectrum of fermionic masses  and required pattern of fermionic mixing.  The radial degree of the flavon decays quickly through its coupling to the SM fermions, but the axial degree can be practically stable to become a DM candidate. 
We have shown that a common solution to DM and the flavor problem of the SM is possible, and can be obtained through a flavonic Goldstone boson in a discrete symmetry framework  accounting for the flavor problem of the SM. 

To achieve this, one needs to introduce a large group leading to a rather high flavor scale,  such as $\mathcal{Z}_{8} \times \mathcal{Z}_{22}$ worked out explicitly in this paper. 
The flavonic dark matter model predicts specific axial flavon coupling to photons which is mostly far below the standard QCD axion DM region, and limited by X-ray searches to $m_\varphi \lesssim 1$ keV and $v_F \gtrsim 4\times 10^{12}$ GeV. 
Thus, there appear no observable consequences in flavour phenomenology.  
Only a limited region of parameter space around $m_\varphi \sim$ neV could be probed by the future radio searches.

It is remarkable that the observed neutrino masses and mixing can be better fitted with Dirac neutrinos,  and thus our framework will be disregarded if neutrinoless double beta decay is found in the forthcoming experiments. 
\begin{acknowledgments} 
EJC and GA are grateful for the support provided by CTP, Jamia Millia Islamia during their visit.  The work of GA is supported by the  Council of Science and Technology,  Govt. of Uttar Pradesh,  India through the  project ``   A new paradigm for flavor problem "  no.   CST/D-1301, and Science and Engineering Research Board, Department of Science and Technology, Government of India through the project `` Higgs Physics within and beyond the Standard Model" no. CRG/2022/003237. 
\end{acknowledgments} 

\appendix 
\section*{Appendix}
    \section*{Benchmark points for the Yukawa couplings}
\label{benchmark}
We use the values of the fermion masses at $ 1$TeV given in ref. \cite{Xing:2007fb}.  The  CKM matrix data are taken from ref.  \cite{Zyla:2021}.  The neutrino  data for the normal hierarchy are used from ref.  \cite{deSalas:2017kay}.   We scan the coefficients $y_{ij}^{u,d,\ell,\nu}= |y_{ij}^{u,d,\ell,\nu}| e^{i \phi_{ij}^{q,\ell,\nu}}$  in the ranges $|y_{ij}^{u,d,\ell, \nu}| \in [0.9,2]$ and $ \phi_{ij}^{q,\ell,\nu} \in [0,2\pi]$.  The results  are,
\begin{equation*}
y^u_{ij} =\begin{pmatrix}
-1.11-0.09 i & 0.15\, +1.50 i  & 0.57\, -0.74 i \\
-1.06 + 0.03 i & -1.30-0.68 i &  -0.95 +0.18 i  \\
1.7\, +0.55 i & 0.68\, +1.63 i & 3.76\, -0.04 i
\end{pmatrix}, 
\end{equation*}

\begin{equation*}
y^d_{ij} = \begin{pmatrix}
0.94\, +0.52 i & 1.27\, +0.79 i & 0.66\, +  i   \\
0.95\, -0.38 i & -0.47 + 0.77 i &  -0.90 + 0.18 i \\
0.92\, +0.01 i & 1.14\, -0.46 i &  0.32\, +1.10 i
\end{pmatrix},  
\end{equation*}
In the standard parametrization, we obtain $\delta_{\rm CP}^q \approx 1.144 = 65.55 \degree$.

\begin{equation*}
y^\ell_{ij} = \begin{pmatrix}
-1.41-0.21 i & 1.14\, -0.008 i & -0.78+0.45i  \\
-0.89 +0.16 i & -0.53 + 0.79 i &  -1.17 +0.10 i \\
0.86\, +0.35 i &  0.91\, +0.003 i & 0.9
\end{pmatrix},
\end{equation*}

For normal mass ordering the neutrino couplings are,
\begin{equation*}
y^\nu_{ij} = \begin{pmatrix}
0.9 & 0.96\, -0.11 i & -0.83 - 0.34 i \\
-1.19+1.61 i & 1.95\, +0.006 i & 1\, -0.22 i \\
0.89\, -1.8 i &  1.13\, +0.1 i & -1.58 + 0.56 i
\end{pmatrix}, 
\end{equation*}
and the leptonic Dirac $CP$ phase  is $\delta_{\rm CP}^\ell  \approx  \pi$.

For inverted mass ordering the neutrino couplings are,
\begin{equation*}
y^\nu_{ij} = \begin{pmatrix}
-2.3-1.13 i& -1.41+2.21 i & -1.08+2.3 i \\
 -0.43-2.9 i & -1.48+0.78 i  & 0.15\, -1.33 i \\
 0.43\, -1.46 i & 0.81\, -0.73 i  & -1.16+1.74 i
\end{pmatrix}, 
\end{equation*}
and the leptonic Dirac $CP$ phase is $\delta_{\rm CP}^\ell  \approx  2.25 = 128.7 \degree$.

\section*{Origin of the  $\mathcal{Z}_{\rm N} \times \mathcal{Z}_{\rm M}$  flavor symmetry}
\label{origin}
 We employ the dark-technicolour (DTC) model  discussed in reference \cite{Abbas:2020frs} to create an origin of the $\mathcal{Z}_{\rm N} \times \mathcal{Z}_{\rm M}$  flavor symmetry.  Let us assume that there are three strong dynamics at a high scale given by the symmetry    $\mathcal{G} \equiv SU(\rm N_{\rm TC}) \times SU(\rm N_{\rm DTC}) \times SU(\rm{N}_{\rm F})$ where TC stands for technicolor,  DTC for dark-technicolor and F represents  a strong dynamics of vector-like fermions.     Moreover,  there are $\rm K_{\rm TC}$ flavors transforming under $\mathcal{G}$ as \cite{Abbas:2020frs},
\begin{eqnarray}
T_{q}^i  &\equiv&   \begin{pmatrix}
T  \\
B
\end{pmatrix}_L:(1,2,0,\rm{N}_{\rm TC},1,1),  \\ \nonumber
T_{R}^i &:& (1,1,1,\text{N}_{\rm{TC}},1,1), B_{R}^i : (1,1,-1,\rm{N}_{\rm TC},1,1),  
\end{eqnarray}
where $i=1,2,3 \cdots$,  and the electric charges $+\frac{1}{2}$ for $T$ and $-\frac{1}{2}$ for $B$.

In a similar manner,  there are $\rm K_{\rm DTC}$ flavors of the $SU(\rm N_{\rm DTC}) $ symmetry  transforming under $\mathcal{G}$ as \cite{Abbas:2020frs},
\begin{eqnarray}
 \mathcal{D}_{ q}^i &\equiv& \mathcal{C}_{L,R}^i  : (1,1, 1,1,\text{N}_{\text{DTC}},1),~\mathcal{S}_{L,R}^i  : (1,1,-1,1,\text{N}_{\text{DTC}},1), 
\end{eqnarray}
where  $i=1,2,3 \cdots$,  and  electric charges $+\frac{1}{2}$ for $\mathcal C$ and $-\frac{1}{2}$ for $\mathcal S$.

The symmetry $SU(N_{\text{F}})$ have the $\rm K_{\rm F}$  fermionic flavors transforming under  $\mathcal{G}$ as \cite{Abbas:2020frs},
\begin{eqnarray}
F_{L,R} &\equiv &U_{L,R}^i \equiv  (3,1,\dfrac{4}{3},1,1,\text{N}_\text{F}),
D_{L,R}^{i} \equiv   (3,1,-\dfrac{2}{3},1,1,\text{N}_\text{F}),  \\ \nonumber 
N_{L,R}^i &\equiv&   (1,1,0,1,1,\text{N}_\text{F}), 
E_{L,R}^{i} \equiv   (1,1,-2,1,1,\text{N}_\text{F}),
\end{eqnarray}
where  $i=1,2,3 \cdots$. In the next step, we assume that there exists an extended-technicolor symmetry whose gauge sector is the mediator among TC, DTC and F fermions.  



In this model there are three axial  $U(1)_{\rm A}$  symmetries, namely,   $U(1)_{\rm A}^{\rm TC, DTC, F }$.   These  symmetries are broken by the  instantons of the corresponding strong dynamics resulting a VEV for  the $2 \rm K_{\rm TC, DTC, F}$-fermion operators,  which does not have any other quantum number such as color or flavor \cite{Harari:1981bs}.  That is,
\begin{equation}
   U(1)_{\rm A}^{\rm TC, DTC, F } \rightarrow \mathcal{Z}_{2 \rm K_{\rm TC, DTC, F}},
\end{equation}
where $\rm K_{\rm TC, DTC, F}$ are number of massless flavors in the fundamental representation of the gauge group $SU(N)_{\rm TC, DTC, F}$.  This breaking results in the conserved axial quantum numbers modulo $2 \rm K$ \cite{Harari:1981bs}.    Therefore, in our theory there are  $\mathcal{Z}_{\rm N} \times \mathcal{Z}_{\rm M} \times \mathcal{Z}_{\rm P}$  residual discrete symmetries where $\rm N= 2 K_{\rm TC}$,  $\rm M= 2 K_{\rm DTC}$, and $\rm P= 2 K_{\rm F}$.  The flavor symmetry  $\mathcal{Z}_{\rm 8} \times \mathcal{Z}_{\rm 22}$ can be obtained by choosing $K_{\rm TC} =4$, i.e.,   four  TC  flavors (2 TC doublets),  and  $K_{\rm DTC}= 11$ DTC flavors.   The  VEV of the flavon field $\chi$ may be  a chiral condensate of the form $\langle \mathcal{D}_L \mathcal{D}_R \rangle$  which further breaks the $\mathcal{Z}_{\rm 8} \times \mathcal{Z}_{\rm 22}$ symmetry. The strong dynamics $SU(\rm N_{\rm F})$ acts like a bridge between the TC and the DTC sectors \cite{Abbas:2020frs}. We note that this UV completion is only for the even discrete symmetry groups. However,  $\mathcal{Z}_{\rm N} \times \mathcal{Z}_{\rm M}$  flavor symmetry may also have some other dynamical origin such as discussed in reference \cite{AristizabalSierra:2014irc}.

\end{document}